\documentclass[sigconf]{acmart}

\copyrightyear{2024}
\acmYear{2024}
\setcopyright{acmlicensed}
\acmConference[ICSE '24]{2024 IEEE/ACM 46th International Conference on
Software Engineering}{April 14--20, 2024}{Lisbon, Portugal}
\acmBooktitle{2024 IEEE/ACM 46th International Conference on Software
Engineering (ICSE '24), April 14--20, 2024, Lisbon,
Portugal}
\acmDOI{10.1145/3597503.3639580}
\makeatletter
\def\@copyrightpermission{This is the author's version of the work. It is posted here for your personal use. Not for redistribution. The definitive version was published in ICSE'24.}
\makeatother

\usepackage[nameinlink,capitalize]{cleveref}
\usepackage{subcaption}
\usepackage{listings}

\lstdefinestyle{default}{
  basicstyle=\ttfamily\footnotesize,
  breaklines=false,
  frame=single,
  framesep=3pt,
  xleftmargin=0pt,
  xrightmargin=0pt,
  prebreak=\raisebox{0ex}[0ex][0ex]{\ensuremath{\hookleftarrow}},
  numbers=none,
  showtabs=false,
  showstringspaces=false,
  tabsize=2,
  numberstyle=\scriptsize,
}
\newcommand{\code}[1]{{\texttt{#1}}}

\begin{document}

\title{Data-Driven Evidence-Based Syntactic Sugar Design}

\author{David OBrien}
\affiliation{
	\institution{Dept. of Computer Science\\Iowa State University}
	\city{Ames}
	\state{IA}
	\country{USA}}
\email{davidob@iastate.edu}

\author{Robert Dyer}
\affiliation{
	\institution{University of Nebraska-Lincoln}
	\city{Lincoln}
	\state{NE}
	\country{USA}}
\email{rdyer@unl.edu}

\author{Tien N. Nguyen}
\affiliation{
	\institution{Computer Science Department\\University of Texas at Dallas}
	\city{Dallas}
	\state{Texas}
	\country{USA}}
\email{tien.n.nguyen@utdallas.edu}

\author{Hridesh Rajan}
\affiliation{
	\institution{Dept. of Computer Science\\Iowa State University}
	\city{Ames}
	\state{IA}
	\country{USA}}
\email{hridesh@iastate.edu}

\newcommand{\edit}[1]{\textcolor{blue}{#1}}


\begin{abstract}
Programming languages are essential tools for developers, and their evolution plays a crucial role in supporting the activities of developers. One instance of programming language evolution is the introduction of syntactic sugars, which are additional syntax elements that provide alternative, more readable code constructs. However, the process of designing and evolving a programming language has traditionally been guided by anecdotal experiences and intuition. Recent advances in tools and methodologies for mining open-source repositories have enabled developers to make data-driven software engineering decisions. In light of this, this paper proposes an approach for motivating data-driven programming evolution by applying frequent subgraph mining techniques to a large dataset of 166,827,154 open-source Java methods. The dataset is mined by generalizing Java control-flow graphs to capture broad programming language usages and instances of duplication. Frequent subgraphs are then extracted to identify potentially impactful opportunities for new syntactic sugars. Our diverse results demonstrate the benefits of the proposed technique by identifying new syntactic sugars involving a variety of programming constructs that could be implemented in Java, thus simplifying frequent code idioms. This approach can potentially provide valuable insights for Java language designers, and serve as a proof-of-concept for data-driven programming language design and evolution.

\end{abstract}

\ccsdesc[500]{Software and its engineering~Language features}

\keywords{syntactic sugars, data-driven language design, subgraph mining}


\maketitle

\section{Introduction}
\label{sec:intro}

Throughout a programming language's lifetime, new features to evolve its expressiveness and functionality such as syntactic sugars are introduced from the intuition and anecdotal experiences of its designers. The adoption and success of these improvements vary wildly~\cite{DyerJavaFeatures} based on the needs of the language's community. In this paper, we argue that this traditional avenue of programming language evolution is flawed due to this variability. Instead, programming language evolution should best serve its community by adhering to the common idioms its developers express, and seek to improve the power of these idioms with complimentary evolution. Thus, this paper advocates for the philosophy of data-driven programming language design and evolution.

The closest approach to data-driven programming language design is the practice of previewing new features prior to release, as seen in programming languages such as Java and Python. In these cases, a new language feature is unofficially released to gather feedback from its user base before finalizing the official design. However, the initial feature is often still proposed based on intuition.

To address this issue, we leverage a large-scale dataset of open-source repositories to extract the frequent patterns found in code. These frequent patterns thus represent common idioms in which a language's features are utilized, and can be valuable sources of information to guide language evolution. Software engineering research is no stranger to frequent pattern mining. These techniques have been leveraged to mine frequent API elements~\cite{Michail00ICSEAssociation, LiPrMiner}, API pairs~\cite{WasylkowskiAnomalies, EnglerInferringErrors, YangPerracotta, WeimerTemporalErrors}, API usage patterns~\cite{NguyenGrouMiner, Grapacc}, change patterns~\cite{NegaraCodeChanges, DilharaPythonChanges, RefactoringMiner1.0, RefactoringMiner2.0}, and code clone detection~\cite{BasitClones, QianClones}.
Previous works~\cite{AllamanisIdioms, AllamanisLoops, SivaramanWild} have mined idioms in source code with non-parametric Bayesian probabilistic tree substitution grammars. Particularly, \citet{AllamanisLoops} mined loop idioms and recommends the \texttt{Enumerate} operator from one such idiom. However, this is the only enhancement argued for, and its users do not evaluate its design. In this study, we aim to recommend multiple enhancements beyond just loop idioms and to evaluate their design amongst participants with relevant experience. 
Additionally, these previous approaches~\cite{AllamanisIdioms, AllamanisLoops, SivaramanWild} are non-deterministic by nature of probabilistic mining. To provide deterministic results, we instead opt to leverage the scalable information-retrieval-based frequent subgraph mining capabilities provided by the Boa  infrastructure~\cite{BoaICSE}.


\begin{figure}[t]
  \begin{subfigure}[t]{0.35\linewidth}
\begin{lstlisting}
int result;
if (n > 0) {
  result = 1;
} else {
  result = -1;
}
\end{lstlisting}
\caption{Not using ternary}
\end{subfigure}
\hspace{6pt}
  \begin{subfigure}[t]{0.5\linewidth}
\begin{lstlisting}
int result = n > 0 ? 1 : -1;
\end{lstlisting}
\caption{Using ternary}
\end{subfigure}
\vspace{-6pt}
    \caption{Example syntactic sugar: {\em ternary operator}}
    \label{ternary}
\end{figure}

This paper aims to mine frequent code patterns that can be simplified by new syntactic sugars to motivate data-driven programming language evolution. As a proof of concept, we focus on the Java programming language, with the goal of recommending new syntactic sugars using techniques to mine Java software repositories at scale~\cite{BoaICSE}. For instance, the {\em ternary operator} is a syntactic sugar depicted in \Cref{ternary}. Suppose a programming language did not support the {\em ternary operator}, but its developers frequently implemented idioms similar to (a) in \Cref{ternary}. The knowledge of this idiom's wide popularity could encourage the implementation of the {\em ternary operator}. Without this knowledge, there might exist missed opportunities for impactful syntactic sugars.

Therefore, to identify frequent idioms not yet supported by an existing syntactic sugar, this paper employs a data-driven approach to analyze frequent control and data patterns in a large corpus of source code. We represent the code as generalized control-flow graphs (CFGs) which model the broad programming language feature usages and models duplicate usage and data patterns. To extract frequent subgraphs from the generalized CFGs, we utilize subgraph mining algorithms provided by Boa \cite{BoaICSE}. These subgraphs are then filtered using a rule-based approach formed from historical observations to identify potential candidates for new syntactic sugars. The resulting set of subgraphs represents common code idioms, which are analyzed and compared with desugared syntactic sugars from other languages. 

In evaluating our generalization, we find that our approach yields a greater quantity of subgraphs with a larger size, frequency, and potential to motivate syntactic sugars than a baseline of mining direct abstract syntax trees which contains no generalization.

To further motivate our approach's effectiveness and the plausibility of data-driven programming language evolution, we organize a catalog of 7 potential new Java syntactic sugars, encompassing multiple programming language functionalities. In total, these syntactic sugars can express common code idioms found millions of times in our corpus. We perform a user study to evaluate the design of our syntactic sugars.

The main contributions of our work are as follows:

\begin{enumerate}
    \item A proof-of-concept for data-driven programming language design and evolution through the recommendation of syntactic sugars simplifying several programming language functionalities.
    \item A generalized control-flow graph representation for broader programming language usage and duplication extraction.
    \item An empirical evaluation against a baseline, highlighting our generalization's effectiveness.
    \item A catalog of 7 potential new Java syntactic sugars that can simplify millions of instances in open-source repositories.
    \item A user study to estimate the user evaluation of the syntactic sugars' design.
\end{enumerate}

This paper is organized as follows: \Cref{sec:motivation} motivates our approach by displaying a catalog of 7 syntactic sugars inspired from idioms mined via our approach. \Cref{sec:approach} explains our approach in detail. \Cref{sec:eval} presents the results of our empirical evaluation, comparing our approach to a sampled baseline. \Cref{sec:results} shows the results of our study on the entire dataset, details the frequencies of our catalog of syntactic sugars, and provides the results of a survey conducted with experienced Java programmers. \Cref{sec:threats} identifies potential threats to the validity of our study. \Cref{sec:related} reviews related works in the area. Finally, \Cref{sec:conclusion} concludes the paper and highlights our main contributions.
\section{Motivation for Syntactic Sugars}
\label{sec:motivation}

Syntactic sugars are expressive and can represent an assortment of idioms through more human-readable or condensed syntax. To motivate data-driven programming language evolution, this section proposes syntactic sugars designed to ``sweeten'' an assortment of frequent idioms. By leveraging our approach, we present 7 new Java syntactic sugars that only took one week to investigate and discuss their designs. These syntactic sugars are diverse, simplifying a set of idioms consisting of a variety of constructs including repeated statements, if-statements, null handlers, and error involvement.

It is important to note that although we propose concrete syntax for these mined sugars, the primary focus of this paper is on identifying frequently occurring idioms that could motivate effective syntactic sugars. There can be many syntactic sugar designs for the same idiom, each with different trade-offs. Thus, the reader is encouraged to focus more on the patterns being identified and less on the actual syntaxes proposed.

\subsection{Repeated Statements}
\label{subsec:repeated}

First, let us consider the case where idioms consist only of the same kind of repeated statement.  Such repetition could possibly be simplified by condensing it with a new syntactic sugar.  As an example, consider the idiom consisting of multiple, successive assignment statements as shown on the left of \Cref{tuple-assignment}.

\begin{figure}[ht!]
\begin{subfigure}[t]{.25\columnwidth}\begin{lstlisting}
id = 0;
name = "Bob";
age = 50;
\end{lstlisting}\caption{desugared}\end{subfigure}
\hspace{8pt}
\begin{subfigure}[t]{.45\columnwidth}\begin{lstlisting}
id,name,age = 0,"Bob",50;
\end{lstlisting}\caption{sugared}\end{subfigure}
    \caption{Potential syntax for Java {\em multiple assignment}}
    \label{tuple-assignment}
\end{figure}

From mining thousands of repositories, our approach finds that this idiom is frequently implemented by Java developers.  We call this the {\em multiple assignments} idiom and propose a Java syntactic sugar for it on the right side of \Cref{tuple-assignment}. Our approach aided in identifying the recurring pattern as a possibility for a new syntactic sugar.  Afterward, we referred to other languages, e.g. Python's tuple unpacking, for inspiration in suggesting the syntax for the new sugar. This sugar can provide a potentially more human-readable syntax to express this Java idiom.

\begin{figure}[ht!]
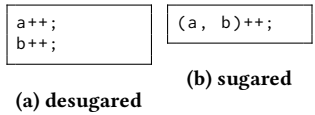

\begin{subfigure}[t]{.2\columnwidth}\begin{lstlisting}
a++;
b++;
\end{lstlisting}\caption{desugared}\end{subfigure}
\hspace{8pt}
\begin{subfigure}[t]{.20\columnwidth}\begin{lstlisting}
(a, b)++;
\end{lstlisting}\caption{sugared}\end{subfigure}
    \caption{Potential syntax for {\em Java multiple \code{++}}}
    \label{multiple-++}
\end{figure}

As another example, consider when multiple increment operators (++) appear in succession, such as on the left of \Cref{multiple-++}.  Our approach found this to be another frequently occurring idiom.  To address this repetitive and potentially redundant code, we propose a new syntactic sugar, as shown on the right of \Cref{multiple-++}. This sugar allows developers to perform multiple increments with a single, concise operation. This sugar can provide syntax which is potentially easier to read and write for Java developers.

\subsection{Negation of \code{If}-Statement}

In addition to repeated statements, we mined several idioms involving \code{if}-statements.  Here we present potential syntactic sugars motivated by our mined idioms. Instead of eliminating duplication like the sugars previously discussed, the sugars in this category aim to improve human readability by rephrasing the \code{if}-statements to better communicate their intended logic. The implementation of these syntactic sugars can potentially improve the readability and understandability of these frequent idioms.

\begin{figure}[ht!]
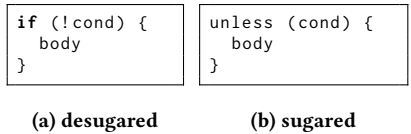

\begin{subfigure}[t]{.25\columnwidth}\begin{lstlisting}
if (!cond) {
  body
}
\end{lstlisting}\caption{desugared}\end{subfigure}
\hspace{8pt}
\begin{subfigure}[t]{.3\columnwidth}\begin{lstlisting}
unless (cond) {
  body
}
\end{lstlisting}\caption{sugared}\end{subfigure}
    \caption{Potential syntax for Java \code{{\em unless}}}
    \label{unless}
\end{figure}

Following the mining of over 166 million method bodies, we find a frequent idiom involving the negation of an \code{if}-statement's condition. Taking inspiration from programming languages like Perl and Ruby which support an {\em unless} statement to invert an if-statement, we propose the introduction of such a construct in Java. \Cref{unless} portrays the design of the proposed \code{unless}-statement.

\begin{figure}[ht!]
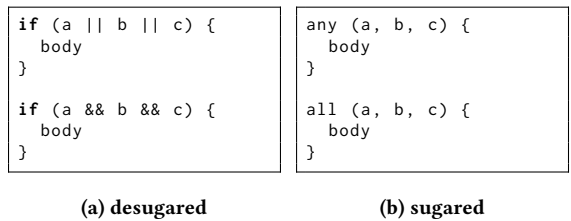

\begin{subfigure}[t]{.40\columnwidth}\begin{lstlisting}
if (a || b || c) {
  body
}

if (a && b && c) {
  body
}
\end{lstlisting}\caption{desugared}\end{subfigure}
\hspace{8pt}
\begin{subfigure}[t]{.40\columnwidth}\begin{lstlisting}
any (a, b, c) {
  body
}

all (a, b, c) {
  body
}
\end{lstlisting}\caption{sugared}\end{subfigure}
    \caption{Potential syntax for \code{{\em Java any/all}}}
    \label{anyall}
\end{figure}

Additionally, \code{if}-statements with repeated conditional operators (\texttt{\&\&} and \texttt{||}) are a frequent idiom, motivating the potential impact of a construct to better express them. \Cref{anyall} exemplifies a separate syntactic sugar of {\em any} and {\em all} syntactic sugars to express \texttt{||} and \texttt{\&\&} conditions respectively. These sugars can potentially viewed as easier to understand, despite requiring the same number of tokens to express as the original idiom.

\subsection{\code{Null} Handlers}

\begin{figure}[ht!]
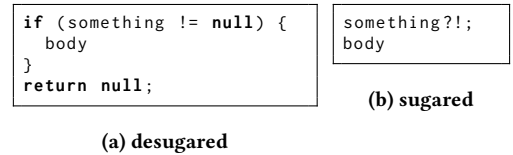

\begin{subfigure}[t]{.45\columnwidth}\begin{lstlisting}
if (something != null) {
  body
}
return null;
\end{lstlisting}\caption{desugared}\end{subfigure}
\hspace{8pt}
\begin{subfigure}[t]{.25\columnwidth}\begin{lstlisting}
something?!;
body
\end{lstlisting}\caption{sugared}\end{subfigure}
    \caption{Potential syntax for Java \emph{null if null}}
    \label{nullifnull}
\end{figure}

Alongside previously discussed categories, our approach also extracts frequent idioms that handle \code{null} values. Therefore, frequent null-handling operations provide an impactful opportunity for simplification.
In this section, we propose a Java {\em null if null (?!)} syntactic sugar to express the behavior shown in a mined frequent idiom shown on the left side of \Cref{nullifnull}. 

The proposed {\em null if null} operator (\texttt{?!}) is similar to Kotlin's {\em not-null assertion} operator (\texttt{!!}) but behaves differently. The proposed operator skips the body if the respective variable is \code{null} and always returns null if no other value is returned in the body. This could compress this frequent handling of null values in Java similar to how syntactic sugars of other languages provide null handlers such as null conditionals and null coalesces.

\subsection{Error Involvement}

Handling errors is a frequent task performed by developers in many situations. Therefore, this category separates itself from previous sections by exploring syntactic sugars designed to express patterns of error-handling code such as throwing or catching errors.

\begin{figure}[ht!]
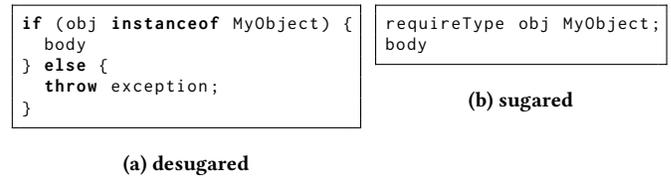

\begin{subfigure}[t]{.52\columnwidth}\begin{lstlisting}
if (obj instanceof MyObject) {
  body
} else {
  throw exception;
}
\end{lstlisting}\caption{desugared}\end{subfigure}
\hspace{8pt}
\begin{subfigure}[t]{.43\columnwidth}\begin{lstlisting}
requireType obj MyObject;
body
\end{lstlisting}\caption{sugared}\end{subfigure}
    \caption{Potential syntax for Java \code{{\em requireType}}}
    \label{typeAssert}
\end{figure}

A frequently recurring idiom found in our mining process is one that involves checking if a variable is an instance of a provided type and triggering an error if it does not meet this requirement. The left side of \Cref{typeAssert} exemplifies such an instance.

To address the verbosity of this common pattern and convey its desired behavior, we propose the \code{{\em requireType}} syntactic sugar and exemplify its usage in \Cref{typeAssert}. This operator has the potential to effectively reduce the amount of code required to express such a frequently implemented idiom.

\begin{figure}[ht!]
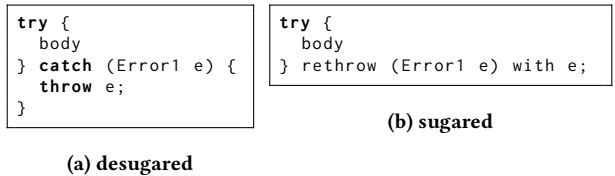

\begin{subfigure}[t]{.36\columnwidth}\begin{lstlisting}
try {
  body
} catch (Error1 e) {
  throw e;
}
\end{lstlisting}
\caption{desugared}
\end{subfigure}
\hspace{8pt}
\begin{subfigure}[t]{.51\columnwidth}\begin{lstlisting}
try {
  body
} rethrow (Error1 e) with e;
\end{lstlisting}
\caption{sugared}
\end{subfigure}
    \caption{Potential syntax for Java \code{{\em rethrow}}}
    \label{errortoerror}
\end{figure}

Continuing our examination, we find another syntactic sugar that can be motivated by a frequently occurring error-handling idiom. This second idiom is a catch block dedicated to rethrowing another error. This can assist when developers want to catch broad exception types, but omit catching specific ones.

\Cref{errortoerror} demonstrates an instance of the mined idiom and the proposed syntactic sugar's design. Accompanying the previously discussed \code{{\em requireType}}, the proposed \code{{\em rethrow}} syntactic sugar offers an alternative for simplifying error-catching and rethrowing operations by specifying which error is to be caught and which is to be thrown in response. Both syntactic sugars presented have the potential to advance the expression of error-handling idioms.

Although the sugars presented here are all new features, our approach can also motivate language features currently being developed by Java designers, such as the upcoming {\em string templates} being previewed in Java 21~\cite{StringTemplateSite}. Our approach mined idioms that involve the composition of string literals, which could have prompted designing a feature such as {\em string templates}.

\section{Approach}
\label{sec:approach}

\begin{figure*}[t]
    
    \centering
    \includegraphics[width=\linewidth]{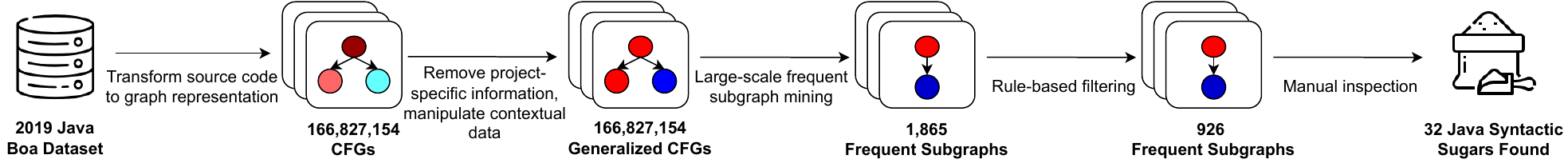}
    \vspace{-12pt}
    \caption{Overview of the workflow of our approach for discovering data-driven syntactic sugars }
    \label{ddss-overview}
\end{figure*}

In this section, we describe our approach for extracting potential syntactic sugars from a large corpus of Java source code. \Cref{subsec:background} provides background on frequent subgraph mining. \Cref{subsec:dataset} discusses the datasets mined. \Cref{subsec:generalized-cfgs} describes our approaches for mining control-flow graphs. \Cref{subsec:frequentsubgraphs} explains how we filter and manually analyze potential syntactic sugar subgraphs. An overview of our approach is depicted in \Cref{ddss-overview}.

\subsection{Background}
\label{subsec:background}

In the context of a graph database, frequent subgraph mining \cite{Apriori1, Apriori2, Apriori3, patterngrowth1, patterngrowth2, significant1, significant2} is the process of identifying patterns and structures that appear frequently across a set of graphs. The task involves extracting subgraphs, which are subsets of the nodes and edges of a given graph, that appear in a number of separate graphs of the input database at least as many times as a user-defined minimum frequency threshold value. The subgraphs are then grouped based on their similarity, with two subgraphs being considered the same if they are isomorphic to each other. This means that the same subset of node and edge types are connected in the same way in separate graphs in the database. The frequency of a subgraph is determined by the percentage of graphs in which it appears, which must be greater than the user-defined minimum frequency threshold in order for it to be considered a frequent subgraph. This approach allows for the identification of common patterns, which can be used for further analysis and understanding of the underlying structure of the entire graph set.

\begin{figure}[ht]
    \centering
    \includegraphics[width=\linewidth]{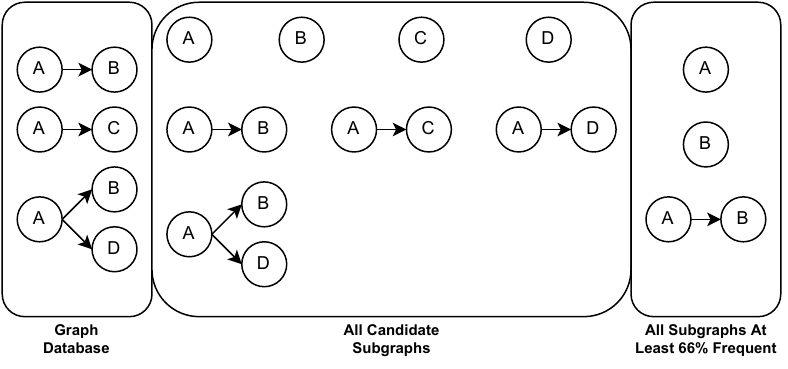}
    \caption{Frequent subgraph mining example of a given graph database and user-defined threshold of 66\%}
    \label{frequent-subgraph-example}
\end{figure}

We provide an example of a frequent subgraph mining task in \Cref{frequent-subgraph-example} where the graph database consists of 3 connected and directed graphs with a set of four node types (A, B, C, and D) and a user-defined minimum frequency threshold of 66\%. To be considered frequent, a subgraph must appear in at least 2 out of the 3 graphs in the database. The graph database is depicted in the leftmost box, while the center box illustrates the set of all possible subgraphs as candidates for being frequent. It is important to note that in frequent subgraph mining, even the subgraph containing each individual node with no edges is considered a candidate subgraph, as well as the subgraph consisting of every node and edge. Finally, the rightmost box shows all subgraphs that have met the minimum frequency threshold of 66\% and appeared in at least 2 out of the 3 graphs in the original database.

In this paper, we leverage the Boa infrastructure because of its proven capabilities in other works~\cite{BoaICSE, DyerJavaFeatures, obrien-prompt-engineering}. Specifically, we use its provided capabilities to mine frequent subgraphs. Boa uses a scalable and deterministic candidate generation approach to gather and aggregate frequent subgraph candidates using Hadoop MapReduce, which enables the efficient handling of large graph sets.

\subsection{Dataset}
\label{subsec:dataset}

In this paper, we leverage a large-scale public dataset provided in Boa. We choose to use this dataset since it is the largest dataset provided, excludes forked projects, and contains frequent subgraph mining capabilities. The dataset statistics are reported in \Cref{oct2019stats}.

\begin{table}[htbp]
  \centering
  \caption{``2019 October/GitHub'' (full) Java dataset statistics}
    \vspace{-0.1in}
    \begin{tabular}{lr}
    \textbf{Granularity} & \multicolumn{1}{r}{\textbf{Amount}} \\
    \toprule
    Projects & 380,125 \\
    Revisions & 23,229,406 \\
    Unique Files & 146,398,339 \\
    File Snapshots & 484,947,086 \\
    AST Nodes & 71,810,106,868 \\
    Recent Snapshot Methods & 166,827,154 \\
    \bottomrule
    \end{tabular}%
  \label{oct2019stats}%
\end{table}%

\begin{figure*}[t]
    \centering
    \includegraphics[width=0.7\linewidth]{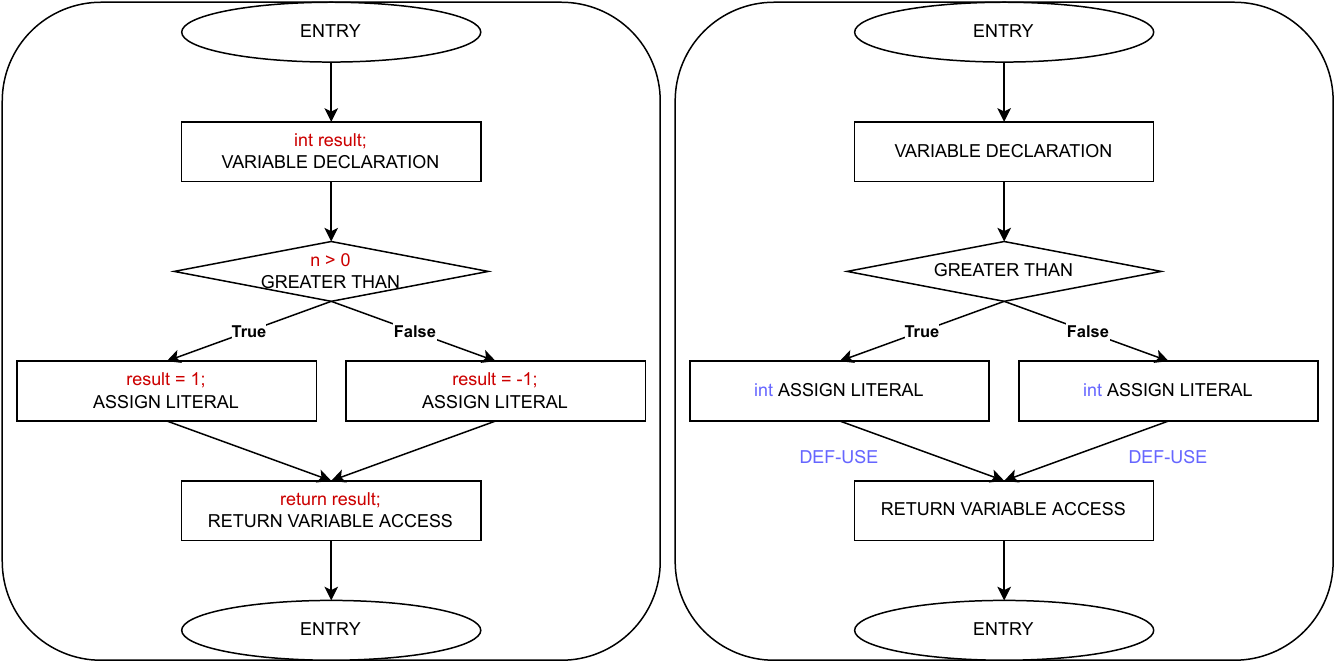}
    \vspace{-0.in}
    \caption{Example control-flow graph (left) and the same control-flow graph with our generalization criteria (right). Information that is dropped from traditional CFG nodes to generalized CFG nodes is shown in red. The information gained in a CFG node during generalization is shown in blue.}
    \label{CFG-Example}
\end{figure*}

We choose to mine the most recent snapshots of all Java methods in our dataset, totaling over 166 million methods. These methods are transformed to CFGs, which is a data structure that has been leveraged in previous studies~\cite{ZhangCodeQA, AcharyaAPIPatterns, NguyenPrecondition, McMillanExemplar, YamaguchiCodePropGraphs}. We utilize Boa as a starting point to modify and analyze these CFGs, which provides capabilities for traversing and mining the information present.

\subsection{Generalized Control Flow Graph}
\label{subsec:generalized-cfgs}

Our goal in this paper is to identify opportunities for new syntactic sugars in a programming language by analyzing the current usage of language features to discover opportunities to advance the syntax. Previous work~\cite{AllamanisIdioms} found that performing frequent subgraph mining on an unmodified tree structure of source code results in very small and meaningless idioms. This is because only small and meaningless subgraphs are frequent across multiple projects when all available information is involved in the mining process. To alleviate this, we propose a novel modification of control-flow graphs (CFG) designed for later frequent subgraph mining. CFGs are chosen to mine due to being a popular higher-level representation of source code. The mining of these modified CFGs enables the discovery of frequent code idioms and redundancies that could be simplified with the introduction of new syntactic sugars. We term these simplifiable subgraphs as being ``sugarable''. This approach aims to identify sugarable subgraphs and promote language evolution.

We identify that syntactic sugars implemented in Java previously can be motivated by popular combinations of operations (e.g., {\em +=}) as well as compressing duplicate code (e.g., {\em multiple variable declarations}). Therefore, the subgraphs representing common idioms should express the broad operations and provide the context of duplication between neighboring expressions. During preliminary experimentation, we found that the broad similarities between CFG nodes contain noise that differentiates them from each other. This would cause them to be considered two different subgraphs during our later mining procedures. However, there is contextual information that certain subgraphs might require to contain enough information to confidently be considered as sugarable. Our technique generalizes the CFG nodes such that two CFG nodes that are broadly similar get mapped to the generalized same type to enable frequent subgraph mining of common idioms.


Therefore, our approach manipulates source code data across the CFG nodes and edges. Specifically, our generalization 1) removes project-specific information, 2) re-uses data of other CFG nodes to provide additional context, and 3) represents generalized neighboring data duplication across the CFG edges.

\Cref{CFG-Example} illustrates a CFG without generalization and a CFG with generalization for comparison and is discussed throughout the rest of this subsection.

Our technique innovates atop traditional CFGs by mapping similar programming language features through discarding specific information (e.g., variable names, user-defined types, and literal values other than \code{null}) as depicted in red in \Cref{CFG-Example}. Discarding project-specific information is crucial in allowing similar subgraphs to be treated as the same during the frequent subgraph mining process. Discarding irrelevant information to increase analysis performance is an action also explored by~\cite{UpadhyayaAccel, AllenStaged, ChoiSparseDataFlow, SmaragdakisSetBased}, but no works consider which pieces of information are relevant in the context of sugarable subgraph mining. As a result of our novel application of this process, frequent code idioms can be identified and considered frequent regardless of project-specific information. 

For instance, in \Cref{CFG-Example} two integer assignments (\code{result = 1} and \code{result = -1}) are originally represented as separate CFG nodes in the traditional CFG due to the difference in literal values getting assigned. However, these integer assignment nodes are considered the same type of node in our generalized representation after the removal of project-specific information from the original representation (variable name and assigned value). This allows for the detection of patterns such as the desugared usage of the {\em ternary operator} from this specific subgraph. This broad pattern would have been difficult to uncover through frequent subgraph mining with the original CFG representation which contained project-specific and subgraph-differentiating information. Therefore, such a subgraph is sugarable because of the duplicate operators in a control-flow structure and previous languages supplying a syntactic sugar atop this idiom. If this subgraph is found to be a frequently-occurring idiom, this would motivate the potential impact of supporting the {\em ternary operator} if it had not already been implemented in Java. Our process of removing project-specific information and generalizing CFG nodes thus helps identify sugarable subgraphs.

In \Cref{CFG-Example}, the information introduced to new areas of a CFG during our generalization process is depicted in blue text. Our technique balances the need for generalization with the importance of context by reapplying information evident in other locations of a CFG. For instance, the original CFG representation involving \texttt{result}'s assignment in \Cref{CFG-Example} lacks the assigned variable's type. However, our approach tracks this information from the variable declarations earlier in the CFG and reapplies it to generalized nodes whenever the variable is assigned a value, providing additional context for understanding potential type-specific syntactic sugars (e.g., {\em string interpolation}) which could simplify the assignment. Without this information, the generalized CFG nodes would simply be \texttt{ASSIGN LITERAL}, lacking potentially crucial details about the type of literal being assigned and introducing ambiguity which can prevent a subgraph from being viewed as sugarable. Thus, our approach strikes a balance between generalization and the provision of the context of extracted subgraphs.

\begin{figure}[t]
    \centering
    \includegraphics[width=\linewidth]{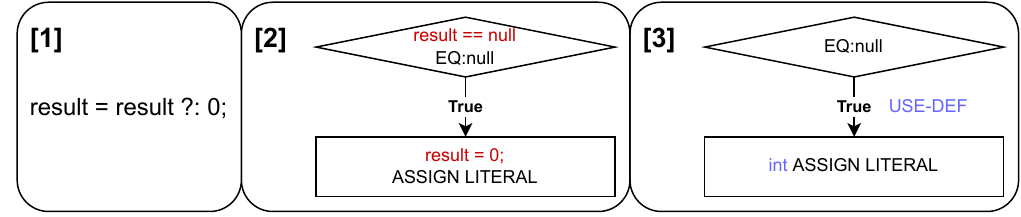}
    \vspace{-12pt}
    \caption{$[1]$ The {\em Elvis operator} in Kotlin,\\ $[2]$ Java equivalent CFG subgraph of desugared {\em Elvis operator},\\ $[3]$ our approach's generalization of $[2]$}
    \label{elvis-example}
\end{figure}

Our approach extends the traditional CFG representation by generalizing neighboring variable definitions and usages. Conventionally, data dependence analysis forms edges to depict the dependence of a variable's usage on its latest definition. However, this does not provide spatial information and can lead to edges connecting nodes far apart in the original source code. In this study, we aim to extract idioms consisting of neighboring CFG nodes to be considered as syntactic sugar opportunities. An example of a sugarable idiom consisting of neighboring information is the {\em Elvis operator (?:)} in Kotlin, where the syntactic sugar implicitly checks if a variable's value is null, and if so, evaluates to a second specified value. Usage of this syntactic sugar in an assignment statement is shown in \Cref{elvis-example}. When considering the desugared equivalent of the {\em Elvis operator}, it's crucial to know that the same variable is both used (checked against null) and redefined (assigned to provided value) in separate expressions. This highlights how syntactic sugars can simplify \textit{neighboring} code idioms that involve the same data, which is not captured by traditional data dependence. 

To address this, we modify the neighboring edges in the CFG to carry additional data context. Our approach introduces four types of edge modifiers: \texttt{DEF-DEF}, \texttt{DEF-USE}, \texttt{USE-DEF}, and \texttt{USE-USE}. These edge modifiers represent whether the source node defines or uses a variable defined or used in the destination node. By incorporating this generalized information, our approach captures important local relationships between expressions, making it easier to identify and motivate syntactic sugars with the applied context while still generalizing these cases to enable large-scale subgraph mining. One such data edge, \texttt{USE-DEF}, is shown in our generalized CFG of Kotlin's {\em Elvis operator}'s desugared Java equivalent in \Cref{elvis-example}.

\subsection{Frequent Subgraph Mining}
\label{subsec:frequentsubgraphs}

A weakness of frequent subgraph mining is the dependence on the user-defined threshold to determine whether a subgraph appears in enough graphs to be considered frequent. Therefore, we choose to leverage a different programming language's established syntactic sugars to mine a threshold of potential syntactic sugar use cases: Kotlin, a programming language that compiles into the Java Virtual Machine (JVM) was created as an alternative to Java when creating Android applications and contains syntactic sugars that Java does not. We identify 4 syntactic sugars in Kotlin that do not have a Java equivalent ({\em string interpolation}, {\em Elvis operator}, {\em getter and setter properties}, and {\em not-null assertion}). Since Java does not contain these syntactic sugars, we consider what their desugared equivalents, the expanded form of the syntactic sugar, would appear as in Java. Following this, our Java dataset is mined for how often these desugared Kotlin syntactic sugars appear. Thus, any subgraph that appears at least as often as the least frequently appearing desugared syntactic sugar from Kotlin is viewed as a candidate for syntactic sugar.

Although mining Kotlin syntactic sugars may bias our results towards Kotlin-like syntactic sugars as opposed to Java-like syntactic sugars, the four chosen syntactic sugars, despite being selected from Kotlin, are not exclusive to Kotlin and are also present in multiple programming languages. For example, {\em string interpolation}, one of the selected syntactic sugars, is not unique to Kotlin and can also be found in other languages such as Python, Scala, and Swift, among others. Similarly, {\em getter and setter properties}, another selected sugar, are present in C\#. Additionally, the {\em not-null assertion} appears in TypeScript, and null-handling syntactic sugars similar to Kotlin's {\em Elvis operator} can be found in C\# and Ruby, albeit under different names. In doing so, although the syntactic sugars were chosen from a similar language to Java, we aim to reduce bias by mining those syntactic sugars in Kotlin which are language agnostic.

\begin{table}[htbp]
  \centering
  \caption{Occurrences of desugared Kotlin syntactic sugars found in 2019/Java Boa dataset}
  \vspace{-6pt}
    \begin{tabular}{lrr}
    \textbf{Syntactic Sugar} & \multicolumn{1}{l}{\textbf{Java Method \#}} & \multicolumn{1}{l}{\textbf{Java Method \%}} \\
    \toprule
    String Interpolation & 1,992,642 & 1.194\% \\
    Elvis Operator & 391,010 & 0.234\% \\
    Getter/Setter Properties & 8,721,251 & 5.228\% \\
    Not-Null Assertion & 101,675 & 0.061\% \\
    \midrule
    \textbf{Total:} & \textbf{166,827,154} & \textbf{100}\% \\
    \bottomrule
    \end{tabular}%
  \label{kotlin-sugars-found}%
\end{table}%

The results of our experiment are shown in \Cref{kotlin-sugars-found}, with {\em not-null assertion} being the least frequent, appearing in 0.061\% of methods. We use this as our threshold for frequent subgraph mining since any code idiom that appears more often would be at least as impactful on existing code as implementing the {\em not-null assertion} syntactic sugar from Kotlin. Although there might be other desugared syntactic sugars not mined that appear less often than the {\em not-null assertion}, using a lower threshold would only add more subgraphs to our results, but all subgraphs presented in this paper would still be considered frequent.

\subsubsection*{Filtering process} Following the extraction of frequent subgraphs from our generalized CFGs, we employ a filtering process to efficiently investigate the potentially large number of subgraphs. We establish a set of rules, drawing inspiration from previous updates to the Java programming language. By basing these rules on previous enhancements, we hope to uncover new sugars that can align with Java's previous evolutionary patterns.

Specifically, we define the following rules to filter the extracted frequent subgraphs.

\begin{enumerate}

    \item \textbf{Duplication:} syntactic sugars such as {\em multiple variable declaration} involve compressing multiple CFG nodes of the same type. This rule captures subgraphs where all nodes are of the same type.
    \item \textbf{Data Edge:} because sugars such as the {\em ternary operator} involve data reuse between nodes, this rule captures all subgraphs with at least one data edge.
    \item \textbf{Null:} Java 8 introduced the \texttt{Optional} class to enable users to assist in operating upon null values. This rule captures subgraphs where at least one node involves a null literal.
    \item \textbf{Error Handling:} syntactic sugars such as {\em multiple catches} involve sugars to simplify error handling idioms. This rule captures subgraphs where at least one node is a "Try", "Catch", or "Throw" node kind.
    \item \textbf{Entry and Exit:} Java 16 provides \texttt{Records} to simplify the creation of data classes and generate certain methods entirely. This rule captures subgraphs that contain both the "Entry" and "Exit" nodes.

\end{enumerate}

The labeling process involved in this study is as follows. For a collection of frequent subgraphs, the subgraphs are first divided by the number of nodes contained. Each subgraph of size 1 is then manually inspected by the first author. Although subgraphs of size 1 might be seen as small-scale, it is important to note that these are CFG subgraphs of size 1, which can contain multiple potentially simplifiable AST nodes. The investigated CFG subgraph is noted if there is enough information present to allow a syntactic sugar to simplify. When evaluating a subgraph for syntactic sugar, the authors consider whether the idiom represented by the subgraph is a complete entity or just a piece of a larger construct (e.g., the beginning of an if-statement rather than the entire body). If the subgraph is not a self-contained unit, it may require additional contextual information to determine if a syntactic sugar is relevant to the situation. Additionally, subgraphs with many nodes and edges that introduce ambiguity may not be suitable for syntactic sugar. Another criterion considered is whether the introduction of syntactic sugar can lead to code compression, reduced duplication, or enable a more expressive operation. The authors also consider whether the subgraph shares similarities with existing syntactic sugars in another programming language. If a known syntactic sugar already exists that addresses a similar coding pattern or idiom, its relevance to the subgraph is considered before deciding if the subgraph motivates a similar syntactic sugar.

After discussion between the authors, there existed frequent subgraphs that the authors believed motivated a syntactic sugar, but were unsure of a proper syntax to propose. These cases are considered sugarable in our evaluation, although a named syntactic sugar is not assigned. However, to provide transparency, all frequent subgraphs as well as their assigned labels are available in our replication package. Following the completion of frequent subgraphs of size $n$, we continue labeling all frequent subgraphs of size $n + 1$ which pass one of the listed filtering rules until no new named syntactic sugars are motivated which were not already motivated by frequent subgraphs of a smaller size.

\section{Empirical Evaluation}
\label{sec:eval}

In this section, we evaluate our mining approach to extract multiple, frequent, and sugarable frequent subgraphs.

To evaluate the effectiveness of our approach, we establish a baseline for comparison. We apply the same frequent subgraph mining and labeling techniques described in Section 3 on a dataset of Java control flow graphs (CFGs) that have not undergone any generalization process. This dataset consists of pretty-printed Java abstract syntax tree (AST) nodes extracted from Boa~\cite{BoaICSE}. By comparing the results of this evaluation to our generalization criteria, we aim to demonstrate that our approach of extracting syntactic sugars from open-source repositories through generalization is more effective than directly mining the AST nodes containing all available information.

\begin{table}[htbp]
  \centering
  \caption{Sampled ``2019 October/GitHub'' Java Boa dataset}
    \vspace{-0.1in}
    \begin{tabular}{lr}
    \textbf{Granularity} & \multicolumn{1}{l}{\textbf{Amount}} \\
    \toprule
    Projects & 7,988 \\
    Revisions & 31,645 \\
    Unique Files & 191,945 \\
    File Snapshots & 622,613 \\
    AST Nodes & 92,311,223 \\
    Recent Snapshot Methods & 241,264 \\
    \bottomrule
    \end{tabular}%
  \label{oct2019smallstats}%
\end{table}%

However, due to the large-scale nature of our original dataset, which comprises over 166 million Java methods, using Boa's frequent subgraph mining algorithms on CFG node types derived from pretty-printed AST nodes resulted in insufficient memory errors. This is because CFGs without generalization have a large number of potential candidate subgraphs. To overcome this limitation, we evaluate a dataset sampled from the original Java Boa dataset, with the dataset statistics provided in \Cref{oct2019smallstats}.

From this sampled dataset, we extracted the subgraphs with our generalization scheme described in \cref{sec:approach} as well as extracted the subgraphs of the pretty-printed Java AST nodes. For each set of extracted subgraphs, we followed the labeling strategy described in \cref{sec:approach}, where we increasingly label all subgraphs of increasing size until no new syntactic sugars can be motivated.

For each set of extracted subgraphs, we use the following metrics:

\begin{enumerate}

\item \textbf{Total frequent subgraphs}: The number of frequent subgraphs that were extracted at each size.
\item \textbf{Investigated subgraphs}: The number of frequent subgraphs that were manually examined after being filtered according to the criteria outlined in \cref{sec:approach}.
\item \textbf{Median frequency}: The median frequency of all frequent subgraphs at the current size. This represents the median number of CFGs in which the frequent subgraphs appear.
\item \textbf{Sugarable}: The number of frequent subgraphs that were labeled as sugarable after the labeling process was complete.
\item \textbf{New sugars}: The number of frequent subgraphs that motivate a new syntactic sugar that was not previously motivated by a frequent subgraph of a smaller size.
\item \textbf{Unique sugars}: Of all the sugarable frequent subgraphs, the number of uniquely named sugars that are involved with the frequent subgraphs at this size. This metric is different from "new sugars" because it includes both new and previously motivated sugars at each size interval, while "new sugars" only include new sugars at that specific size.

\end{enumerate}

\begin{table}[htbp]
  \centering
  \caption{Results for sampled Boa Dataset {\em with generalization}}
  \vspace{-6pt}
    \begin{tabular}{lrrrrr}
          & \multicolumn{1}{|p{1cm}|}{\centering \textbf{Total Freq. SGs}} & \multicolumn{1}{|p{1cm}|}{\centering \textbf{Invest. SGs}} & \multicolumn{1}{|p{1cm}|}{\centering \textbf{Median Freq.}} & \multicolumn{1}{|p{1cm}|}{\centering \textbf{Sugar-\\able}} & \multicolumn{1}{|p{1cm}|}{\centering \textbf{New/ \\ Unique Sugars}} \\
          \toprule
    1 node  & 156   & 156   & 814   &  44   & 13/13 \\
    2 nodes & 628   & 249   & 337   & 116   & 16/16 \\
    3 nodes & 592   & 309   & 294   &  57   & 4/11 \\
    4 nodes & 377   & 173   & 260   &  11   & 0/6 \\
    \bottomrule
    \end{tabular}%
  \label{sample_generalization}%
\end{table}%

\begin{table}[htbp]
  \centering
  \caption{Results for sampled Boa Dataset {\em without generalization} (pretty-printed Java AST)}
  \vspace{-6pt}
    \begin{tabular}{lrrrrr}
          & \multicolumn{1}{|p{1cm}|}{\centering \textbf{Total Freq. SGs}} & \multicolumn{1}{|p{1cm}|}{\centering \textbf{Invest. SGs}} & \multicolumn{1}{|p{1cm}|}{\centering \textbf{Median Freq.}} & \multicolumn{1}{|p{1cm}|}{\centering \textbf{Sugar-\\able}} & \multicolumn{1}{|p{1cm}|}{\centering \textbf{New/ \\ Unique Sugars}} \\
          \toprule
    1 node  & 148   & 148   & 205   & 8   & 3/3 \\
    2 nodes & 106   &  22   & 214   & 0   & 0/0 \\
    \bottomrule
    \end{tabular}%
  \label{sample_prettyprint}%
\end{table}%

The results for the generalized frequent subgraphs and the pretty-printed Java AST subgraphs are presented in \Cref{sample_generalization,sample_prettyprint}, respectively. From the data, we draw the following conclusions:

\begin{enumerate}

\item \textbf{Subgraph size:} By using our generalization criteria, we were able to identify sugarable frequent subgraphs that consisted of four CFG nodes, as opposed to only finding sugarable subgraphs with one CFG node without generalization. This demonstrates that our generalization criteria enable the discovery of sugarable subgraphs that are larger than our established baseline, i.e., without generalization.

\item \textbf{Number of subgraphs:} Across all investigated subgraph sizes, our approach extracted more frequent subgraphs when using our generalization criteria as opposed to not using generalization. This is because our generalization approach allows for CFG nodes with similar use cases to be considered the same CFG node type, rather than only considering two CFG nodes as the same type if they are identical. This leads to the discovery of more frequent subgraphs, and thus more potential areas to discover sugarable programming patterns.

\item \textbf{Subgraph frequency:} Across all investigated subgraph sizes, our approach with generalization found subgraphs that were more frequent than the baseline. This means that the subgraphs investigated are likely to have a larger impact than the syntactic sugars extracted from frequent subgraphs mined without generalization.

\item \textbf{Syntactic sugars discovered:} Across all investigated subgraphs, our approach discovered more sugarable subgraphs and more new syntactic sugars not previously found. This is because our generalization criteria capture the broad programming language use cases, allowing for common patterns to appear as frequent subgraphs instead of infrequent when mining subgraphs without any generalization.

\end{enumerate}

From our empirical evaluation, we find that our generalization criteria in frequent subgraph mining lead to a significant improvement in the extraction of sugarable results. The results obtained from the sampled data demonstrate that the use of generalization results in the identification of larger, more abundant, and frequent subgraphs, as well as an increase in the number of sugarable subgraphs. The results of our evaluation demonstrate that our generalization criteria significantly enhance the performance of frequent subgraph mining, highlighting the potential of this method in uncovering more meaningful and useful programming patterns.
a\section{Empirical Results}
\label{sec:results}


\subsection{Sugarable Pattern Discovery Result}

This section presents the results of our technique applied to the complete dataset of 166 million Java source code methods. These results have been obtained through a manual examination of the data and are available in the replication package~\cite{replication-package}. We evaluated the entire corpus of 166,827,154 CFGs available in the ``2019 October/GitHub'' Boa dataset with the same evaluation metrics outlined in \Cref{sec:approach}. The findings are presented in \Cref{entire_generalization}. We provide an analysis of the effectiveness of our technique on a large scale.

Running our tool on the full Java dataset, we were able to extract a total of 1,865 frequent subgraphs, which range in size from 1-4 nodes. Through the application of our rule-based filtering method outlined in \Cref{sec:approach}, 926 of these subgraphs were selected for further manual investigation. While a significant portion of the initial set of frequent subgraphs was removed through this filtering process, the goal of this study is to identify new opportunities for syntactic sugars in the Java programming language. To that end, the rules used in the filtering process were specifically designed to target subgraphs that align with previous updates in Java.

The results of our manual examination of the filtered frequent subgraphs yielded a total of 241 sugarable subgraphs. We consider a subgraph to be sugarable if it exhibits characteristics such as redundancies that can be compressed, operations that can be combined, or if it resembles a known syntactic sugar from another programming language. It is worth noting that some sugarable subgraphs were not assigned a specific named syntactic sugar due to the lack of confidence in the appropriate syntax to simplify the respective subgraph. Overall, our proposed approach aimed to identify opportunities for new syntactic sugars that can simplify Java code and improve its readability and maintainability.

In our results, we identified {\em 32 named syntactic sugars} that have the potential to simplify the frequent subgraphs that we have identified. It is important to note that this number is considered to be a lower bound, as more experienced programming language designers may be able to discover additional syntactic sugars that were not identified through our manual investigation process. The full dataset is available in our replication package~\cite{replication-package} so that others can inspect and identify additional sugars. In order to provide a clear understanding of our results, the 7 syntactic sugars presented previously in \Cref{sec:motivation} were selected from amongst these 32 ones.

\begin{table}[t] 
  \centering
  \caption{Results for entire Boa Dataset with generalization}
  \vspace{-6pt}
    \begin{tabular}{lrrrrr}
          & \multicolumn{1}{|p{1cm}|}{\centering \textbf{Total Freq. SGs}} & \multicolumn{1}{|p{1cm}|}{\centering \textbf{Invest. SGs}} & \multicolumn{1}{|p{1cm}|}{\centering \textbf{Median Freq.}} & \multicolumn{1}{|p{1cm}|}{\centering \textbf{Sugar-\\able}} & \multicolumn{1}{|p{1cm}|}{\centering \textbf{New/ \\ Unique Sugars}} \\
          \toprule
    1 node  & 163 & 163 & 522,965 &  50 & 13/13 \\
    2 nodes & 669 & 246 & 238,973 & 120 & 16/16 \\
    3 nodes & 639 & 324 & 205,421 &  59 & 3/10 \\
    4 nodes & 394 & 193 & 200,701 &  12 & 0/6\\
    \bottomrule
    \end{tabular}%
  \label{entire_generalization}%
\end{table}%

\subsection{Frequency Results}

\begin{table}[htbp]
  \centering
  \caption{Amount of CFGs for different syntactic sugars}
  \begin{tabular}{lr}
    \toprule
    \textbf{Sugar} & \textbf{Amount of CFGs} \\
    \midrule
    Multiple Assignment & 5,543,853 \\
    Multiple \code{++} & 292,449 \\
    \code{Unless} & 9,574,658 \\
    \code{Any/All} & 10,267,105 \\
    {\em Null if Null} & 236,567 \\
    \code{requireType} & 116,742 \\
    \code{Rethrow} & 1,994,698 \\
    \bottomrule
  \end{tabular}
  \label{amount-of-cfgs}
\end{table}

In \Cref{sec:motivation}, a catalog of syntactic sugars was presented to preview the results of our technique. We now present the frequencies of all the subgraphs that motivate these syntactic sugars' creation in \Cref{amount-of-cfgs}. It is worth noting that these numbers serve as a lower bound, as there could be infrequent subgraphs that also motivate the same syntactic sugar. Additionally, this sum only represents the number of CFGs that contain at least one instance of the subgraph. By nature of frequent subgraph mining, if multiple instances of a frequent subgraph appear in a single CFG, it is not multiply counted. By presenting the frequency of the subgraphs that motivate each syntactic sugar, we demonstrate the potential impact that could be achieved by incorporating these syntactic sugars into future versions of Java. These results serve as a motivator for language designers to adopt data-driven programming language evolution. 

\subsection{Survey on Human Subjects}

Adoption of new language features can vary~\cite{KimYiSugar, DyerJavaFeatures} and syntactic sugar specifically can introduce ambiguities in code understanding~\cite{GopsteinUnderstanding}. Therefore, to evaluate the quality of the designed sugars presented in \Cref{sec:motivation}, we conducted a survey involving 31 participants with experience in Java programming. In the survey, participants are asked to provide their title and their years of experience with Java. The results are reported in \Cref{participant-titles} and \Cref{participant-year}. As seen, our participants consist of both graduate students and practitioners. The majority of our participants have 5 or more years of experience. Participants are then presented with the desugared and sugared figures shown in \Cref{sec:motivation} and asked for their preference on a Likert scale. The results of these questions are depicted in \Cref{participants-results}.

\begin{table}[t] 
  \centering
  \caption{survey participants}
    \begin{tabular}{lr}
          \toprule
           & \textbf{Frequency} \\
          \midrule
          Graduate Student & 48.39\% \\
          Practitioner & 45.16\% \\
          Other & 6.45\% \\
          \bottomrule
    \end{tabular}%
  \label{participant-titles}%
\end{table}%

\begin{table}[t]
  \centering
  \caption{Experience in Java of survey participants} 
    \begin{tabular}{lr}
          \toprule
          \textbf{Experience} & \textbf{Frequency} \\
          \midrule
          0-2 years & 16.13\% \\
          2-5 years & 22.58\% \\
          5-10 years & 38.71\% \\
          10-15 years & 6.45\% \\
          15-20 years & 3.23\% \\
          20+ years & 12.90\% \\
          \bottomrule
    \end{tabular}%
  \label{participant-year}%
\end{table}%

\begin{table*}[htbp]
  \centering
  \caption{Results on Participant Preferences \& Feedback on Syntactic Sugars}
    \begin{tabular}{p{3cm}|c|p{11cm}}
          \hline
          \textbf{Sugar} & \textbf{Preference} & \textbf{Comments} \\
          \hline
          \textbf{Multiple Assignment} & 58.06\% & Have the flexibility to put meaningful groups of variables together\\
          Multiple \code{++} & 45.16\% & I think it would be a useful feature. It would be more useful if we can increment by any constant together, not just 1, like (a,b)+=2. \\
          \code{Unless} & 41.94\% & I like the idea of unless, but the "else" case seems awkward. Is there a better word than unless? \\
          \textbf{\code{Any \& All}} & 74.19\% & Seems to read 'easier' (matches how we would state the condition in natural language) and unambiguously, so I'm a fan of the new feature. \\
          {\em Null if Null} & 29.03\% & The syntax appears to perform an implicit return in a way that is very subtle and unclear. A null-coalescence operator would be helpful, however. \\
          \code{requireType} & 45.16\% & I would actually go a step further and suggest a more robust pattern matching syntax, similar to rust or scala, which allows exhaustively enumerating types/patterns. \\
          \textbf{\code{Rethrow}} & 51.61\% & This makes it very elegant. A very useful suggestion.
 \\
          \hline
    \end{tabular}%
  \label{participants-results}%
\end{table*}

Our survey reveals that over half of our participants prefer three of our proposed syntactic sugars compared to their desugared equivalents, showcasing the potential of the syntactic sugars discovered. For example, 74.19\% of participants favor \code{{\em ``Any} and {\em All''}} syntactic sugars over the current Java code. Participants have described them as {\em "Seems to read `easier' (matches how we would state the condition in natural language) and unambiguously, so I'm a fan of the new feature."} Another participant notes {\em "the new features would allow for composition that reduces the potential for errors arising from confusion over operator precedence and associativity."} Thus, we conclude that our approach often identifies idioms that can inspire user-preferred syntactic sugars.

It is a possibility that our survey results depict a lower bound, since the participants are asked about the syntactic sugar's design, it might be possible that participants might prefer an alternative syntactic sugar to simplify the same coding idiom. We provide comments where the participants recommend modifications to our controversial syntactic sugar designs in \cref{participants-results}. Therefore, this survey shows for each sugar the minimum preference that users might have for sugaring our identified idioms since they might potentially be more popular sugars to express our mined idioms.

\section{Threats to Validity}
\label{sec:threats}

In our mining process, we utilized the 2019 Java dataset provided by Boa. This dataset was chosen as it contains thousands of diverse projects and provides support for frequent subgraph mining, which is essential for our study. Other large-scale Java datasets could be used in our study. However, note that newer versions of Java have been released since 2019, and new syntactic sugars and language usage patterns may have emerged. Despite this, we conducted a manual inspection of the extracted frequent subgraphs and considered cases where a new Java syntactic sugar or an existing syntactic sugar could simplify the subgraph. These results are available in our replication package. It is also worth noting this is only possible due to the wealth of Java data available. This approach may not apply to less popular programming languages.

It is also worth noting that some of the files in the dataset may be perfect duplicates of each other. Previous research has shown that Java repositories have a low percentage of file clones, with 58\% of the files being distinct~\cite{DejaVuDuplicates}. However, these clones could still impact our results if the presence of duplicated files skewed the frequency of certain subgraphs in our dataset. This is important to keep in mind when interpreting and applying our results.

Throughout this study, decisions were made which may introduce bias into our results. First, we employed a rule-based approach to filter our extracted subgraphs which were formed atop observations made regarding previous Java changes. Additionally, subgraphs were manually investigated by one author. The goal of this study is to motivate the plausibility of data-driven programming language design and evolution through mining frequent code idioms, rather than exhaustively identifying all possible syntactic sugars to implement in Java. We argue that the presentation of 7 new syntactic sugars mined from our approach and the performed survey's results show the effectiveness of our decisions. Additionally, any recommendable sugarable subgraphs that were missed (either filtered out by our rule-based approach or misclassified by manual reviewing) are still available on our replication package for future researchers and language designers to refer to.

\section{Related Work}
\label{sec:related}

Previous works \cite{AllamanisIdioms, AllamanisLoops} utilized probabilistic mining techniques to extract frequent code idioms. In the context of refactorings, Sivaraman {\em et al.}~\cite{SivaramanWild} extended that research; however, these approaches are non-deterministic. In contrast, we adopt a deterministic information-retrieval approach to facilitate data-driven language design and evolution. Notably, this topic was explicitly excluded from the scope of prior research which leverages idiom mining for refactorings \cite{SivaramanWild}. \citet{AllamanisLoops} mines loop idioms to suggest that LINQ could benefit from the \texttt{Enumerate} using a mined loop idiom as evidence. However, our work recommends multiple enhancements beyond just loop idioms, and we also evaluate our designs in a user study consisting of experienced Java programmers.

Mining the patterns of library usages has been a frequently explored application of frequent pattern mining~\cite{Michail00ICSEAssociation, LiPrMiner, WasylkowskiAnomalies, EnglerInferringErrors, YangPerracotta, WeimerTemporalErrors}. \citet{NguyenGrouMiner, Grapacc} proposes techniques to mine and utilize context-preserving graph representations of source code for anomaly detection and code completion. However, these approaches are focused on API usage, not programming language usage, which is the main focus of our study.

In the area of code refactorings, \citet{BritoJavaRefactorings} characterized non-trivial subgraphs representing refactorings in Java and JavaScript applications, while \citet{JankeTSECodeChange} mined version control systems to extract 25 frequent change patterns found across multiple software projects. Additionally, detecting and classifying common code changes has been explored by other works~\cite{NegaraCodeChanges, DilharaPythonChanges, RefactoringMiner1.0, RefactoringMiner2.0}. However, these works focus on change patterns and not patterns of the most recent usage of programming language features.

Our approach of generalizing a graph representation of source code is comparable to works that remove irrelevant elements to produce more efficient and fruitful results~\cite{UpadhyayaAccel, AllenStaged, ChoiSparseDataFlow, SmaragdakisSetBased}. However, prior works do not explore the necessary information to preserve for sugarable subgraph mining.
\section{Conclusion}
\label{sec:conclusion}

In this paper, we proposed a data-driven approach for programming language design and evolution by identifying common code idioms through frequent subgraph mining, and manually evaluating the extracted subgraphs for opportunities to implement impactful syntactic sugars. To accomplish this goal, we have generalized 166,827,154 CFGs to capture the broad programming language usage patterns. From this process, we found 241 total sugarable subgraphs and specifically cataloged and evaluated 7 potential new syntactic sugars for Java, including syntactic sugars involving duplication, if-statements, nulls, and errors. Our empirical results demonstrate the feasibility of data-driven programming language design and evolution, exemplifying new Java syntactic sugars that can simplify millions of common programming idioms.

\section{Data Availability}

The Boa queries, output from those queries, and all processing scripts are made available in a replication package on Zenodo~\cite{replication-package}.

\begin{acks}
Generative AI was leveraged to revise the writing throughout the paper, but no sections were entirely generated by generative AI. We thank Ali Ghanbari for his feedback regarding the initial design of the proposed syntactic sugars. This work is supported by the National Science Foundation under the grants CCF-15-18897, CNS-15-13263, CCF-19-34884, CNS-21-20448, CNS-21-20386, and CCF-22-23812.
\end{acks}

\balance
\bibliographystyle{ACM-Reference-Format}
\bibliography{refs}


\end{document}